# Resolving the Hydrophobicity of Me-4PACz Hole Transport Layer for High Efficiency Inverted Perovskite Solar Cells

Kashimul Hossain,[1,†] Ashish Kulkarni,[2,†]* Urvashi Bothra,[1] Benjamin Klingebiel,[3] Thomas Kirchartz,[3,4] Michael Saliba,[2,5] and Dinesh Kabra[1]*

† contributed equally to this work.

[1]Department of Physics, Indian Institute of Technology Bombay, Powai, Mumbai 400076, India.

[2]Helmholtz Young Investigator Group FRONTRUNNER, IEK5-Photovoltaik, Forschungszentrum Jülich, Wilhelm-Johnen-Straße, 52428 Jülich, Germany

[3]IEK-5 Photovoltaik, Forschungszentrum Jülich, Wilhelm-Johnen-Straße, 52428 Jülich, Germany.

[4]Faculty of Engineering and CENIDE, University of Duisburg-Essen, Carl-Benz-Str. 199, 47057 Duisburg, Germany.

[5]Institute for Photovoltaics (IPV), University of Stuttgart, Pfaffenwaldring 47, 70569 Stuttgart, Germany.

*Corresponding authors – a.kulkarni@fz-juelich.de and ashish.kulkarni786@gmail.com, dkabra@iitb.ac.in

# ABSTRACT

[4-(3,6-Dimethyl-9H-carbazol-9-yl)butyl]phosphonic acid (Me-4PACz) self-assembled monolayer (SAM) has been employed in perovskite single junction and tandem devices demonstrating high efficiencies. However, a uniform perovskite layer does not form due to the hydrophobicity of Me-4PACz. Here, we tackle this challenge by adding a conjugated polyelectrolyte poly(9,9-bis(3'-(N,N-dimethyl)-N-ethylammonium-propyl-2,7-fluorene)-alt-2,7-(9,9dioctylfluorene))dibromide (PFN-Br) to the Me-4PACz in a specific ratio, defined as Pz:PFN. With this mixing engineering strategy of Pz:PFN, the PFN-Br interacts with the A-site cation and is confirmed via solution-state nuclear magnetic resonance studies. The narrow full width at half maxima (FWHM) of diffraction peaks of perovskite film revealed improved crystallization on optimal mixing ratio of Pz:PFN. Interestingly, the mixing of PFN-Br additionally tunes the work function of the Me-4PACz as revealed by Kelvin probe force microscopy and built-in-voltage estimation in solar cells. Devices employing optimized Pz:PFN mixing ratio deliver open-circuit voltage ($V_{OC}$) of 1.16V and efficiency >20% for perovskites with a bandgap of 1.6 eV with high reproducibility and concomitant stability. Considering significant research on Me-4PACz SAM, our work highlights the importance of obtaining a uniform perovskite layer with improved yield and performance.

# INTRODUCTION

The power conversion efficiency (PCE) of perovskite solar cells (PSCs) has rapidly increased from 3.8% to a certified value of > 25%.[1–3] This rapid rise in the PCE can be credited to excellent optoelectronic properties such as a high absorption coefficient, low exciton binding energy, and a tuneable bandgap.[4–8] The so-called inverted or *p-i-n* device architecture is of significant interest using low-temperature processed charge transport layers (CTLs) such as poly(3,4-ethylenedioxythiophene) polystyrene sulfonate (PEDOT: PSS), poly (triaryl amine) (PTAA), fullerene ($C_{60}$) are employed.[9–13] Recently, Al-Ashouri *et al.*, used various phosphonic acid group anchored carbazole-based self-assembled monolayers (SAMs) such as 2PACz, MeO-2PACz, and Me-4PACz as hole transport layers (HTL), demonstrating superior device performance than the widely employed PTAA as HTL.[14–16] In particular, by employing Me-4PACz SAM as HTL, the same group, reported one of the highest efficiency in single junction devices and a certified PCE of ~29% in silicon-perovskite two-terminal tandem solar cells.[16] In comparison to other SAMs, the Me-4PACz-based device is reported to demonstrate the lowest density of interface traps indicating suppressed non-radiative recombination at the perovskite/Me-4PACz interface.[17] These salient features make Me-4PACz SAM a promising HTL for PSCs. Motivated by this, we fabricate PSCs with Me-4PACz as HTL. Unfortunately, we (and others) observed poor perovskite layer formation on the Me-4PACz and is in line with recent reports.[18–21] Strategies such as the incorporation of an $Al_2O_3$ insulating layer at the interface of Me-4PACz and perovskites and deposition of Me-4PACz by evaporation have been reported to improve the perovskite layer coverage on Me-4PACz.[18,19,22] These strategies are promising, but developing a strategy to deposit Me-4PACz by solution process and simultaneously obtain a uniform perovskite layer with improved device performance with a proper understanding remains a challenge.

Until recently, the interface engineering approach has been applied to address the issue of poor wetting of perovskite on underlying hydrophobic polymeric HTLs. For instance, PTAA is known to be a hydrophobic HTL and modification of its surface with poly(9,9-bis(3'-(N,N-dimethyl)-N-ethylammoinium-propyl-2,7-fluorene)-alt-2,7-(9,9dioctylfluorene))dibromide (PFN-Br), poly(methyl methacrylate) (PMMA):2,3,5,6-tetrafluoro-7,7,8,8-tetracyanoquinodimethane (F4-TCNQ), polyvinyl oxide (PEO) and tetra-n-propylammonium bromide (TPAB), phenylethylammonium iodide (PEAI), etc. have been reported.[12,23–27] Among others, modification with PFN-Br has been widely employed as it helps in obtaining a uniform perovskite layer with augmented device efficiency. Previously, we have also employed PFN-Br modified PTAA and reported PCE of >20%.[9] In addition to polymeric HTLs, modification of SAMs has also been reported.[28,29] Recently, Li *et al.*, employed a mixture of MeO-2PACz and 2PACz to improve the charge extraction and reported a PCE of 25% for all-perovskite tandem solar cells.[29] Deng *et al.*, showed that co-assembled monolayers (a mixture of SAM and alkylammonium containing SAM) can help in simultaneously suppressing the non-radiative recombination and surface functionalization and the resultant device showed a PCE of 23.59%.[28] Very recently, Al-Ashouri added 1,6-hexylenediphosphonic acid with Me-4PACz to improve the perovskite layer deposition; however, a detailed investigation has not been performed. Until present, the modification of carbazole-based SAMs has been limited to co-assemble with phosphonic or carboxylic acid anchoring group which might limit the exploration of SAM. Therefore, strategies need to be developed, beyond co-assembly with SAM, to improve the perovskite layer coverage on Me-4PACz SAM.

In this work, we report a mixing engineering strategy by combining Me-4PACz SAM with conjugated polyelectrolyte PFN-Br to obtain a highly reproducible and improved performance of perovskite solar cells. With an aid of mixed Me-4PACz:PFN-Br, a uniform perovskite thin film is obtained and the device with optimized Me-4PACz:PFN-Br mixing ratio showed a PCE of >20% in 0.175 cm$^2$ device active area. To the best of our knowledge, the obtained device efficiency is one of the highest values reported for Me-4PACz with a triple cation perovskite composition having a bandgap ($E_g$) of 1.6 eV. More importantly, with the help of solution nuclear magnetic resonance (NMR), X-ray diffraction (XRD) and Kelvin probe force microscopy (KPFM) outcomes, we elucidate the triple role of PFN-Br. The PFN-Br interacts with the A-site cation and improves the crystallization. This also elevates the valence band position of Me-4PACz, leading to better interfacial energy level alignment with perovskite having a bandgap of 1.6 eV. We note that this improved performance could be achieved due to reduced interfacial localized states (low series resistance ($R_S$)) and bulk defects in the absorber layer (ideality factor $n$ ~1), which result not only in high fill factor (*FF*), short-circuit current density ($J_{SC}$) and open-circuit voltage ($V_{OC}$) but also provide an excellent yield and stability for optimal PSCs. The un-encapsulated device showed enhanced stability by retaining the initial device performance at T$_{95}$ even after 3000 hours when measured in ~ 40% humidity conditions.

## Results and Discussion

Me-4PACz has been used as HTL in a single junction and tandem (with silicon) devices demonstrating one of the highest efficiencies.[16,17,30] In an attempt to fabricate the device using Me-4PACz SAM, we observed poor perovskite layer formation. The schematic illustration of perovskite deposition on the Me-4PACz coated ITO substrate and the photographic image (and a **video V1**) of a perovskite thin film is shown in **Figure 1a**. The poor perovskite layer coverage can be attributed to the presence of non-polar groups such as methyl (–$CH_3$) and long-alkyl chain ($C_4H_8$) in Me-4PACz SAM which are in general responsible for hydrophobic nature.[31–35] To understand this more clearly, we performed water contact angle measurements of Me-4PACz and compared it with widely used hydrophobic polymer and small organic molecules based HTLs such PTAA, poly[2,6-(4,4-bis-(2-ethylhexyl)-4H-cyclopenta [2,1-b;3,4-b′]dithiophene)-alt-4,7(2,1,3-benzothiadiazole)] (PCPDTBT), poly(3-hexylthiophene-2,5-diyl) (P3HT) and 4,4',4"-Tris[phenyl(m-tolyl)amino]triphenylamine (MTDATA). **Figure S1** shows the molecular structure of the aforementioned HTLs and their respective water contact angle in comparison with Me-4PACz. As expected, the water contact angle for Me-4PACz is high and in the same range as compared to other HTLs evidencing its hydrophobic nature. We additionally washed the Me-4PACz coated substrate with methanol (solvent to dissolve SAM; see the experimental section in Supplementary Information), in an attempt to reduce the hydrophobicity and obtain a uniform perovskite film. However, the perovskite layer showed non-uniform coverage, as shown in **Figure S2**. This indicates that the phosphonic acid group strongly binds with ITO and a strong monolayer fingerprint is present even after the washing step and is in line with previous reports.[14] To further confirm the role of non-polar groups that are present in Me-4PACz in preventing the perovskite layer formation, we fabricated device by employing 2PACz SAM as HTL. Because of the absence of the –$CH_3$ group and long alkyl chain, not only a uniform perovskite layer was obtained but the resultant device showed a PCE of ~ 20%, this further confirms that other than the SAM all other layers of the device stack are working as expected. The current density ($J$) – voltage ($V$) curve of the best-performing device employing 2PACz SAM is shown in **Figure S3**. These results evidently suggest that Me-4PACz is sufficiently hydrophobic to prevent the formation of a uniform perovskite layer and strategies need to be developed to improve the perovskite layer coverage on the Me-4PACz SAM.

Typically, researchers have incorporated PFN-Br as an interlayer to modify the interface of hydrophobic polymeric HTLs and obtained a uniform perovskite layer.[23,36] We initially modified the surface of Me-4PACz with PFN-Br and obtained a uniform perovskite layer. The relevant discussion is mentioned in ESI-1 and the results are shown in **Figures S4** and **S5**. The performance of these devices was low and therefore we performed a mixing engineering strategy. The steps involving device stack layers deposition are schematically shown in **Figure 1b**. The Me-4PACz and PFN-Br were mixed in 6:4, 7:3, 8:2, 9:1, and 9.5:0.5 ratios followed by the perovskite layer deposition by one-step method (please see the experimental section for more details). From now onwards, for our convenience, we term Me-4PACz:PFN-Br as Pz:PFN. Irrespective of all the mixing ratios, a uniform perovskite layer was obtained as shown in Figure 1b and **Figure S6**. Top surface scanning electron microscope (SEM) images, see **Figure S7**, showed no significant difference in the perovskite layer morphology with respect to the different Pz:PFN ratios. Moreover, we observed indistinct perovskite layer morphology on both the Me-4PACz/PFN-Br (term as Pz/PFN) (**Figure S7a**) dual layer and Me-4PACz (**Figure S7g**) compared to the mixing ratio cases. This further indicates that the perovskite layer crystallization is not influenced by the PFN-Br mixing or interlayer modification. The X-ray diffraction (XRD) diffractograms, as shown in **Figure S8**, showed no traces of residual lead iodide and no significant difference in the perovskite crystal structure. The combined outcomes of SEM and XRD indicate that perovskite crystallizes similarly in all cases without any significant changes. After the perovskite layer deposition, phenyl-$C_{61}$-butyric acid methyl ester (PCBM) and bathocuproine (BCP) were deposited via the solution process as an electron transport layer (ETL) and buffer layer respectively, and the silver electrode was thermally evaporated to complete the cell. **Figure 2a** depicts the best-performing device *J-V* curves in forward bias under 1 Sun illumination for all the mixed Pz:PFN based perovskite solar cells. The *J-V* curves under forward and reverse scan direction are shown in **Figure S9b** and the device parameters are tabulated in **Table S3**. With a change in mixing ratio, the device performance first increased and then decreased. We note that there is a change in $J_{SC}$ (also integrated $J_{SC}$ from IPCE) with respect to the compositional ratio of HTL, which will be discussed later along with KPFM results. The device obtained from Pz:PFN with a 9:1 ratio showed the highest PCE of 20.67% with a $J_{SC}$ of 22.54 mA/cm$^2$, $V_{OC}$ of 1.16 V, *FF* of 79.12% with slight hysteresis (please see **Table S3**). Note that the obtained efficiency is one of the highest values reported with Me-4PACz based 1.6 eV bandgap PSCs device. **Figure 2b** shows the incident photon to current efficiency (IPCE) spectrum of our best-performing device along with the reflection (R), and transmission (T) spectrum. The IPCE spectra of the devices based on different Pz:PFN ratios are shown in **Figure S10**. The dip in the EQE spectrum correlates well with the hump in the reflection spectra. The loss in the current density due to reflection can be calculated from the external quantum efficiency (EQE) and internal quantum efficiency (IQE) spectrum using equations S-E3 and S-E4. The integrated $J_{SC, EQE}$ is calculated by integrating the EQE spectrum over the 1-sun spectrum. There is a current density mismatch by less than 5% of the $J_{SC, EQE}$ to $J_{SC}$ measured from the *J-V* measurement and this can be understood by edge effect from the active area (17.5 mm$^2$ **Figure S9a**) of the device or pre-bias measurement condition.[37,38] To verify the reproducibility of the device performance with Pz:PFN mixing engineering strategy, 30 devices of each mixing ratio were fabricated using the device procedure outlined in the experimental section (in the supplementary information). **Figure 2(c)** summarizes the distribution of photovoltaic parameters under forward scan. The average performance improved from ~17% for Pz:PFN (6:4) to >20% for Pz:PFN (9:1) ratio case with the most narrow distribution among 30 PSCs, i.e., highly reproducible efficient PSCs. The performance improvement is attributed to the increase in $J_{SC}$, $V_{OC,}$ and *FF*. The stabilized efficiency under maximum power point tracking for the Pz:PFN (9:1) PSC device is 20.14% (at 970 mV) inset **Figure 2(a)** ( and for Pz:PFN (6:4) PSC device is 17.32% (at 920 mV) **Figure S9 (c)** )

Numerous research reports have employed PFN-Br as an interlayer to overcome the hydrophobicity of polymeric HTLs such as PTAA to obtain a uniform perovskite film and even as an additive in the perovskite solution to improve its performance.[23,36,39] On the other hand, by looking at the structure of PFN-Br (**Figure S11a**), one can see the presence of a non-polar long alkyl chain. This can cause severe wetting issues and therefore raises similar concerns about forming a non-uniform perovskite layer. Moreover, the high water contact angle of PFN-Br, as shown in **Figure S11b**, corroborates the hypothesis. However, a uniform perovskite layer was formed after the incorporation of PFN-Br. This raises the point of how the perovskite layer is formed when a highly hydrophobic PFN-Br is introduced. In literature, the role of PFN-Br in helping to form a uniform perovskite layer has not been focused on yet. For instance, Wang *et al.* incorporated PFN-Br interlayer in between the PTAA HTL and perovskite absorber layer and assumed the possibility of PFN-Br redissolution during the deposition of the perovskite layer, thereby improvement in the device performance.[39] Therefore, it is of utmost importance to understand the underlying mechanism of a uniform perovskite layer formation on PFN-Br interlayer or mixed Pz:PFN HTL. We noticed that PFN-Br is readily soluble in DMF and DMSO solvents (**Figure S12**) and therefore, we hypothesized that the presence of PFN-Br helps in tuning the perovskite crystallization. To verify this, we deposited MAPbI$_3$ (without Br$^-$) independently on MeO-2PACz and Pz: PFN-coated glass substrate. The MAPbI$_3$ perovskite film deposited on MeO-2PACz and Pz:PFN (8:2) is indistinctly uniform in **Figure S13**. The XRD diffractograms were carried out for both films as shown in **Figure 3a**. The XRD pattern of MAPbI$_3$ deposited on Pz:PFN HTL showed a peak with lower full width at half maximum (FWHM). A slight shift (~0.0024º) in XRD peak is observed between MAPbI$_3$ film deposited on MeO-2PACz coated glass substrate *vs* MAPbI$_3$ film on Pz:PFN (8:2) coated glass substrate. The reason for this small shift is investigated via spectroscopic technique as the spectroscopic probe is known to be more sensitive than the structural probe. As per previous reports by us and others, micro-structure synchrotron studies were found to be less conclusive than spectroscopic studies, where spectroscopy could provide essential insight to correlate with solar cells performance.[40,41] These samples are tested for PL spectroscopy. The raw PL data showed small peak-shift and FWHM differences. FWHM of MAPI film being broad on MeO-2PACz substrates could be explained based on the FWHM of the XRD peak. However, peak-shift which is found to be red-shifted in raw data, interestingly overlapped once an O.D. (or self-absorption) correction is introduced **Figure 3 (b)** and **Figure S14**.[42] This suggests that bromine containing Pz:PFN is not incorporating the Br$^-$ ion into the absorber layer within the sensitivity limits of used structural and spectroscopic probes. A small peak-shift on the higher degree side in the XRD peak falls near to resolution limit of the instrument, which is 0.0012º.

To further understand the interaction between PFN-Br and perovskite, a series of liquid-state $^1$H nuclear magnetic resonance (NMR) measurements in solution (deuterated DMSO) were performed and the results are shown in **Figure 3c – e** (full-scale data shown in **figure S15**). Initially, an $^1$H NMR spectrum of triple-cation perovskite solution dissolved in deuterated DMSO was recorded (**Figure 3c**), showing a resonance peak at around 9 ppm, which corresponds to protons bound to the nitrogen atoms of formamidinium (HC(NH$_2$)$_2^+$ or FA$^+$) and methylammonium (CH$_3$NH$_3^+$ or MA$^+$). Upon addition of PFN-Br into the perovskite solution in deuterated DMSO, the peak at around 9 ppm splits into two new peaks due to interaction between cationic HC(NH$_2$)$_2^+$ (FA$^+$)/CH$_3$NH$_3^+$ (MA$^+$) and PFN-Br. To further examine the interaction between MA$^+$ and PFN-Br, an additional $^1$H NMR experiment of only MAI without and with PFN-Br was carried out. As shown in **Figure 3d**, no peaks are split and shifted indicating that the peak split in the perovskite solution with PFN-Br addition (Figure 3c) is related to the change in the FA$^+$ cation in the presence of PFN-Br. To further confirm this conclusion, we recorded a $^1$H NMR spectrum of only FAI without and with PFN-Br. As shown in **Figure 3e**, this peak around 9 ppm of FAI solution with PFN-Br additive gets split into two peaks in a way similar to the splitting observed in the perovskite solution with PFN-Br additive. The splitting of $FA^+$ protons into two new signals has also been observed by other research groups and is ascribed to the formation of hydrogen bond complexes of the amidinium moiety of the $FA^+$ cation.[43–46] Considering all NMR results, it is thus proposed that a hydrogen bond between $FA^+$ and PFN-Br is existed, which might help in perovskite layer formation during its spin coating step and crystallization and might also remain present in the final perovskite film and can be beneficial for prolong device stability. In addition to the perovskite layer formation with an improved interfacial property, we hypothesized that the addition of PFN-Br might influence the electronic properties of Me-4PACz. Therefore, we measured the work function of Me-4PACz (without and with PFN-Br different ratios), and perovskite using Kelvin Probe Force Microscopy (KPFM) technique. The work function of the films is determined using gold as a reference and the results are shown in **Figure 3f, Figure S16, and Figure S17**. Though the pristine Me-4PACz shows a better energy level alignment with the perovskite layer (**Figure S17**). However, forming the perovskite layer on pristine Me-4PACs is a challenge. Because of the effect of interface dipoles caused by the electrolyte moiety present in PFN-Br, the work function of Me-4PACz showed dramatic changes with the mixing ratios. For the particular Pz:PFN (9:1) case, the energy level alignment matches closely with the Me-4PACz (without any PFN-Br) and also with the work function of the perovskite layer implying better hole extraction efficacy. Moreover, we measured the ultraviolet photoelectron spectroscopy (UPS) of the above mentioned HTLs and the Fermi levels are well aligned with KPFM results as shown in **Figure S17b**. Further, the built-in potential difference ($V_{bi}$) calculated from the dark current characteristics of diodes shows a clear trend in reduced barrier-voltage for the case of Pz:PFN (9:1) composition as compared to the other mixing ratio composition, as seen in **Figure S18.** We observed a slight relative change in the EQE and corresponding $J_{SC}$ (whether integrated or measured using an AM1.5G solar simulator light source), which can be seen modulating with respect to $V_{bi}$ values (see Figure S18) of diodes. $V_{bi}$ values modulate with respect to different Fermi-levels of ITO/(Pz:PFN::X:Y) layer, as ETL is common in all diodes (Figure S17).[47] This also explains the reason behind the high device efficiency Pz:PFN (9:1) case compared to the other cases despite similar perovskite crystal structure and top surface morphology.

The dark current measurement can also give insights into the recombination at the perovskite and Pz:PFN interface by an estimation of reverse saturation current density $J_0$.[48] Therefore, we measured the dark current of the devices with different ratios of Pz:PFN and observed the lowest dark current density $J_D$ at –0.3 V of $5.28 \times 10^{-6}$ mA/cm$^2$ and the reverse saturation current density $J_0$ of $2.20 \times 10^{-10}$ mA/cm$^2$ for the Pz:PFN (9:1) (**Figure S19a**) whereas a typical $J_0$ value for established Silicon photovoltaics is $10^{-10} \sim 10^{-9}$ mA/cm$^2$.[49,50] The $J_D$ and $J_0$ values of all the devices with different Pz:PFN mixing ratios are tabulated in **Table S5**. The lowest $J_D$ and $J_0$ values for the 9:1 case imply suppressed recombination at the perovskite/ Pz:PFN interface. **Figure 4a** represents the variation in $V_{OC}$ vs. $\log_{10}(J_{sc}/J_0)$, where $J_{SC}$ and $J_0$ are calculated from the illuminated and dark currents. The $\log_{10}(J_{sc}/J_0)$ is higher for the Pz:PFN (9:1) (**Figure S19b**) and the corresponding $V_{OC}$ is higher as per the basic diode equation.[51] Transient photovoltage (TPV) decay measurements were carried out to determine the charge carrier lifetime and to quantify the recombination process.[52,53] The normalized TPV decay profile, fitted with a mono-exponential decay and measured for all the cases is shown in **Figure S20a**. Among all the cases studied, the TPV decay profile of the device with Me-4PACz:PFN-Br (9:1) showed a slow decay and a high perturbed charge carrier lifetime ($\tau$) (please refer to **Table S6**) implying suppressed non-radiative recombination.[54,55] This can be further correlated with the device $V_{OC}$. The $V_{OC}$ of the device is calculated as a function of the perturbed charge carrier lifetime ($\tau$) and high $V_{OC}$ as in the case of 9:1 case (**Figure 4b**); implying the lower recombination. The dependence of $V_{OC}$ on the incident light fluence was investigated for all the Pz: PFN-based devices to get further insight into recombination processes. The incident light is varied from 100 mW/cm$^2$ (1 Sun) to 1 mW/cm$^2$ (0.01 Sun) with a set of neutral density (ND) filters. **Figure S20b (and Figure S21)** plots $V_{OC}$ against intensity in a semi-log scale and slope corresponds to $2.303nkT_cq^{-1}$ where $n$, $k$, $T_c$, and $q$ are ideality factor, Boltzmann constant, room temperature of our device, and elementary charge respectively.[56] The slope of the best-performing device, that is, employing Pz:PFN (9:1) gives an $n$ value of 1.05 (Figure S20b) and is smallest when compared to other mixing ratios of Pz:PFN based devices, implying suppressed non-radiative trap assisted recombination. Ideality factor $n$ being close to 1 suggests that the absorber is free from bulk defects and hence it provides higher $V_{OC}$.[48] Relatively, this higher $V_{OC}$ (free from non-radiative recombination channel) further facilitates charge carrier extraction under short circuit conditions, i.e. $J_{SC}$ for optimal PSCs. The lower value of series resistance (~ 4Ω /sq) calculated from the dark and illuminated current, also indicates higher $FF$ and $V_{OC}$ for Pz:PFN (9:1) as shown in **Figure S20 c, d (and Figure S22)**.[57,58] The device performances are limited by the series resistance. If the devices are free from series resistance, then the $FF$ will be higher and the reverse saturation current will be lower which can be calculated from the Suns $V_{OC}$ measurement as shown in **Figure S23**.[59] The Suns $V_{OC}$ based pseudo *J-V* curve showed more than two orders of magnitude reduction in reverse saturation current density $J_0$ to a value of sub-PA, i.e. 0.78 pA/cm$^2$, **Figure S24 and Table S7**.[60] We measured the electroluminescence quantum efficiency (QE$_{EL}$) of the PSCs and found to be higher for the Pz:PFN(9:1) as shown in **Figure S25**. **Figure 4c** depicts the (QE$_{EL}$) plotted as a function of log$_{10}$ ($J_{sc}/J_0$) and it is higher for the Pz:PFN(9:1). This again indicates that the Pz:PFN(9:1) HTL-based PSC interface is having relatively lesser trap states which also verified from higher TPV lifetime (indicating the presence of lower trap states).[61,62] Finally, we investigated the long-term stability of our best performing Pz:PFN (9:1) based device in comparison with Pz:PFN (6:4). The device performance was measured under relative humidity in a range of 40 – 50% and aged in N$_2$-filled glove box without sealing or encapsulation. **Figure S26** shows the long-term stability of our best-performing device. After >3000 hours of storage, the device retained T$_{95}$ >3000 hours of the maximum device efficiency. This improved device yield and stability can be ascribed not only to the uniform perovskite layer coverage on the Me-4PACz but also to the interfacial interaction of PFN-Br with the perovskite.

To summarize, we present a mixing engineering strategy of combining Me-4PACz SAM with the conjugated polyelectrolyte PFN-Br polymer. The mixing of the PFN-Br with the Me-4PACz, facilitates uniform deposition of the perovskite layer on the top of hydrophobic Me-4PACz HTL. The uniform deposition of the perovskite layer on the top of PFN-Br mixed Me-4PACz happened due to Me-4PACz:PFN-Br interaction with A-site cation which was confirmed via solution-state NMR. This further facilitates an improved crystallization of the perovskite layer, which was confirmed via narrow diffraction in XRD, and narrow PL peaks result. In addition to this, the KPFM result reveals that mixing with PFN-Br tunes the work function of Me-4PACz and for the optimized 9:1 mixing ratio the energy level alignment of Pz:PFN matches well with the perovskite. As a result of this, the perovskite device demonstrates high and reproducible efficiency of over 20% concomitant with high stability for T$_{95}$ >3000 hours. when measured in relative humidity of ~ 40%. We believe that the mixing engineering of SAM Me-4PACz with electrolyte polymer PFN-Br will not only open new doors to tackle hydrophobic SAMs in solution-processable efficient photovoltaic devices but also will allow designing new electrolyte-based polymers and/or small molecules that can be combined with SAMs, thereby quilting a better interface.

## Acknowledgments

This work was supported by the Ministry of New and Renewable Energy India: National Centre for Photovoltaic Research and Education (NCPRE) Phase II. This work was also partially supported by the Indo-Swedish joint project funded by DST-India (DST/INT/SWD/VR/P-20/2019). A.K. and M.S. thanks the Helmholtz Young Investigator group FRONTRUNNER.

## Conflict of Interest

The authors declare no conflict of interest.

## AUTHOR INFORMATION

**Corresponding Authors**

Prof. Dinesh Kabra − Department of Physics, Indian Institute of Technology Bombay, Mumbai 400076, India; orcid.org/0000-0001-5256-1465

Email: dkabra@iitb.ac.in

Dr. Ashish Kulkarni - IEK-5 Photovoltaik, Forschungszentrum Jülich, Wilhelm-Johnen-Straße, 52428 Jülich, Germany.

Email: a.kulkarni@fz-juelich.de and ashish.kulkarni786@gmail.com

**Authors contribution:**

K.H. and A.K. contributed equally to this work.

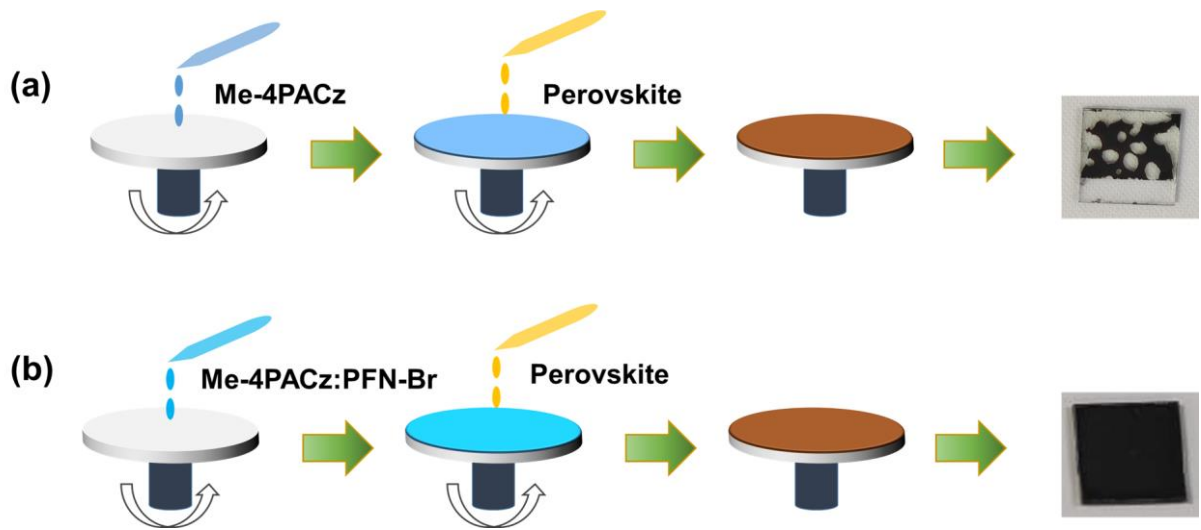

**Figure 1:** Schematic of the thin film deposition of the Me-4PACz, Me:4PACz:PFN-Br (abbreviated as Pz:PFN), and perovskite layer. **(a)** Me-4PACz and perovskite layer deposition. **(b)** Deposition of mixed Me-4PACz:PFN-Br and perovskite layers.

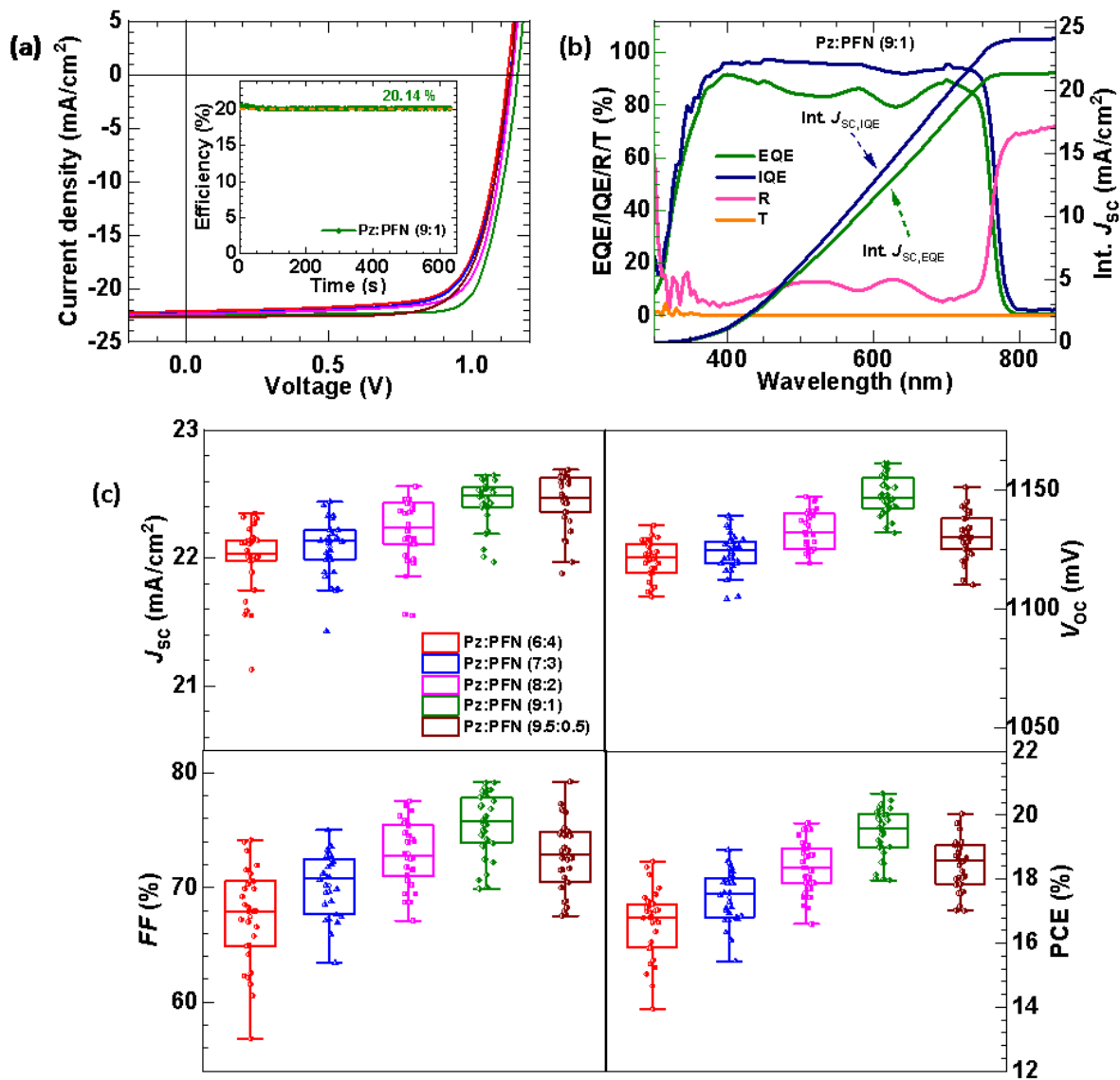

**Figure 2:** (Colour line figures – Pz:PFN (6:4) is red colour, Pz:PFN (7:3) is blue colour, Pz:PFN (8:2) is magenta colour, Pz:PFN (9:1) is olive colour, Pz:PFN (9.5:0.5) is wine colour). **(a)** Current density *vs.* Voltage (*J-V*) characteristics of the photovoltaic devices under 1-sun (100 mW/cm$^2$) condition in the forward scan direction. The forward and reverse *J-V* scans of the representative devices are shown in **Figure S9b**. The inset figure represents the power output of the Pz:PFN (9:1) PSC device taken under maximum power point tracking with stabilized efficiency is 20.14% (at 970 mV). **(b)** IPCE spectrum of the best performed Pz:PFN (9:1) PSC including the reflection (R), transmission (T), and integrated current density (Int. $J_{SC}$). **(c)** Boxplot of short circuit current density ($J_{SC}$), open circuit voltage ($V_{OC}$), fill factor (*FF*), and power conversion efficiency (PCE%) over 30 devices.

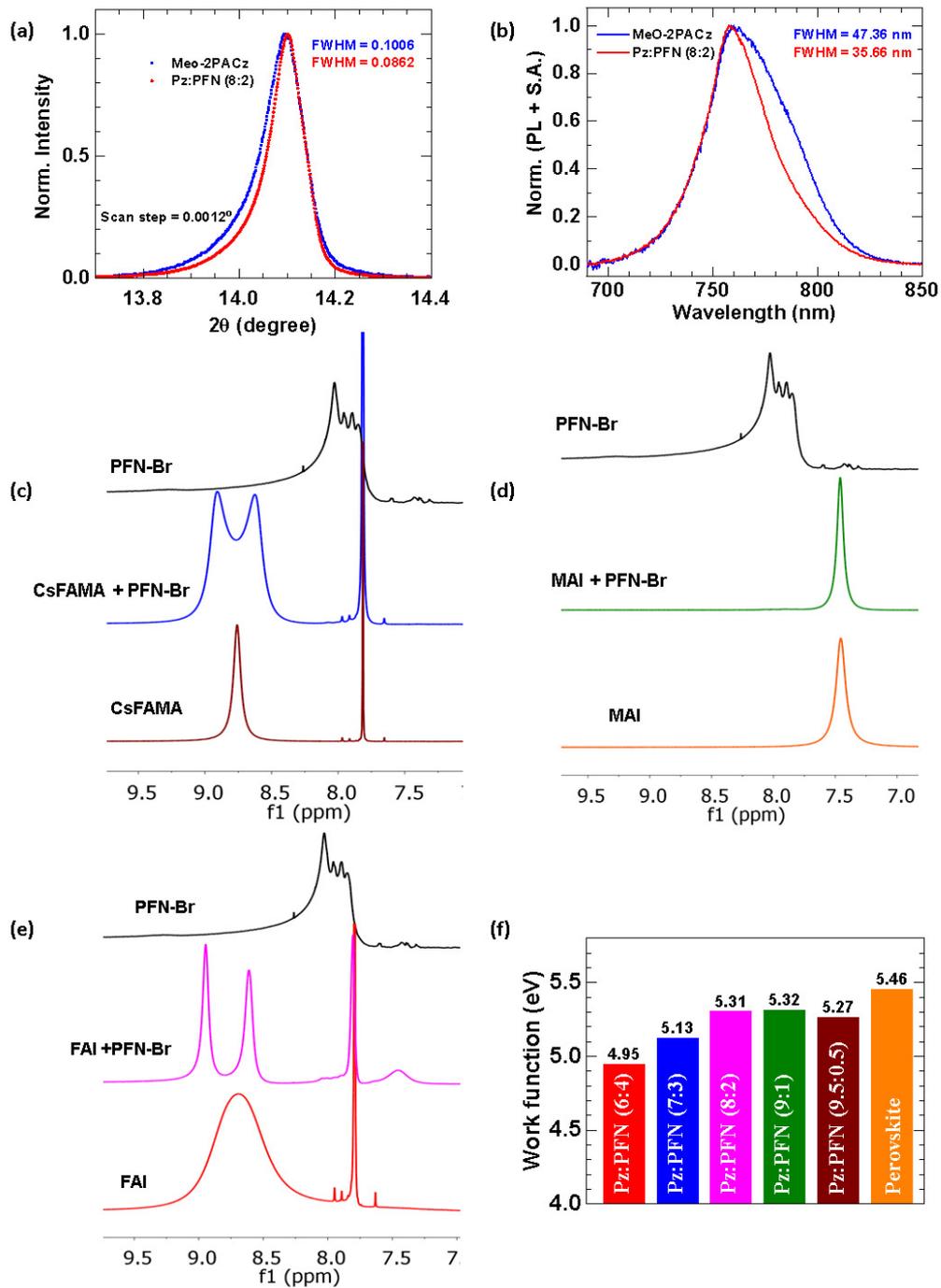

**Figure 3. (a)** XRD pattern and **(b)** PL spectra (self-absorption corrected) of the MAPbI$_3$ perovskite thin film ( thickness ~ 250 nm ) deposited on the MeO-2PACz and Pz:PFN coated glass substrates.[42] **(c)** $^1$H NMR of the CsFAMA, CsFAMA +PFN-Br and PFN-Br solution prepared in DMSO d6 solvent. **(d)** $^1$H NMR of the MAI, MAI +PFN-Br and PFN-Br solution prepared in DMSO d6 solvent. **(e)** $^1$H NMR of the FAI, FAI +PFN-Br, and PFN-Br solution prepared in DMSO d6 solvent. **(f)** The work function of the Pz:PFN HTLs was measured using KPFM study.

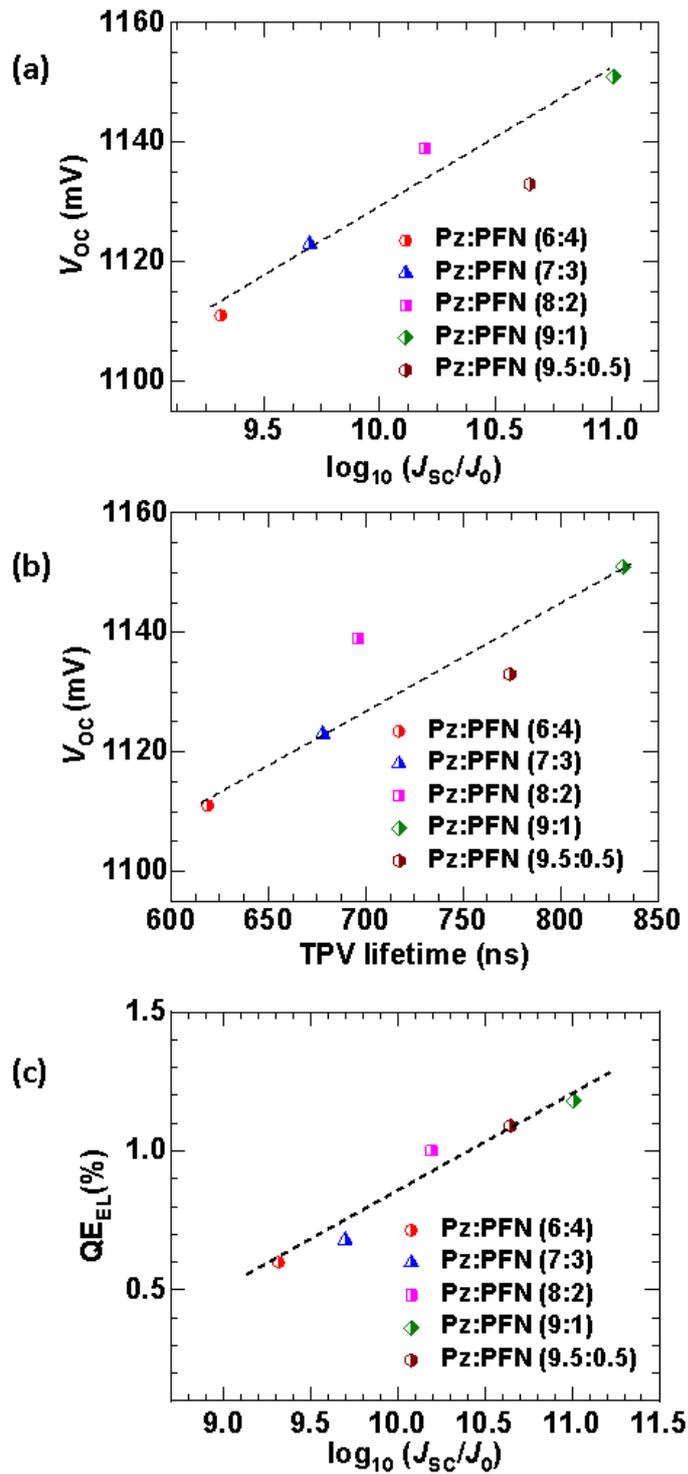

**Figure 4.** **(a)** The $V_{OC}$ of the representative devices as a function of short circuit ($J_{SC}$) and reverse saturation current density ($J_0$). **(b)** The $V_{OC}$ of the representative devices as a function of the perturbed charge carrier lifetime measured from TPV measurement at 1-Sun illumination condition. **(c)** Electroluminescence (EL) quantum efficiency plotted as a function of $J_{SC}$ and $J_0$.

Self-Assembled Monolayers: A Molecular Dynamics Study. *J. Phys. Chem. C* **2020**, *124* (26), 14237–14244.

(32) Godawat, R.; Jamadagni, S. N.; Garde, S. Characterizing Hydrophobicity of Interfaces by Using Cavity Formation, Solute Binding, and Water Correlations. *Proc. Natl. Acad. Sci.* **2009**, *106* (36), 15119–15124.

(33) Gao, Y.; Duan, L.; Guan, S.; Gao, G.; Cheng, Y.; Ren, X.; Wang, Y. The Effect of Hydrophobic Alkyl Chain Length on the Mechanical Properties of Latex Particle Hydrogels. *RSC Adv.* **2017**, *7* (71), 44673–44679.

(34) Chen, W.; Karde, V.; Cheng, T. N. H.; Ramli, S. S.; Heng, J. Y. Y. Surface Hydrophobicity: Effect of Alkyl Chain Length and Network Homogeneity. *Front. Chem. Sci. Eng.* **2021**, *15* (1), 90–98.

(35) Bakulin, A. A.; Pshenichnikov, M. S.; Bakker, H. J.; Petersen, C. Hydrophobic Molecules Slow down the Hydrogen-Bond Dynamics of Water. *J. Phys. Chem. A* **2011**, *115* (10), 1821–1829.

(36) Moot, T.; Patel, J. B.; McAndrews, G.; Wolf, E. J.; Morales, D.; Gould, I. E.; Rosales, B. A.; Boyd, C. C.; Wheeler, L. M.; Parilla, P. A. Temperature Coefficients of Perovskite Photovoltaics for Energy Yield Calculations. *ACS Energy Lett.* **2021**, *6* (5), 2038–2047.

(37) Saliba, M.; Etgar, L. Current Density Mismatch in Perovskite Solar Cells. *ACS Energy Lett.* **2020**, *5* (9), 2886–2888.

(38) Singh, S.; Shourie, R. J.; Kabra, D. Efficient and Thermally Stable CH3NH3PbI3 Based Perovskite Solar Cells with Double Electron and Hole Extraction Layers. *J. Phys. D. Appl. Phys.* **2019**, *52* (25), 255106.

(39) Wang, H.; Song, Y.; Kang, Y.; Dang, S.; Feng, J.; Dong, Q. Reducing Photovoltage Loss at the Anode Contact of Methylammonium-Free Inverted Perovskite Solar Cells by Conjugated Polyelectrolyte Doping. *J. Mater. Chem. A* **2020**, *8* (15), 7309–7316.

(40) Jain, N.; Chandrasekaran, N.; Sadhanala, A.; Friend, R. H.; McNeill, C. R.; Kabra, D. Interfacial Disorder in Efficient Polymer Solar Cells: The Impact of Donor Molecular
21

**TOC**

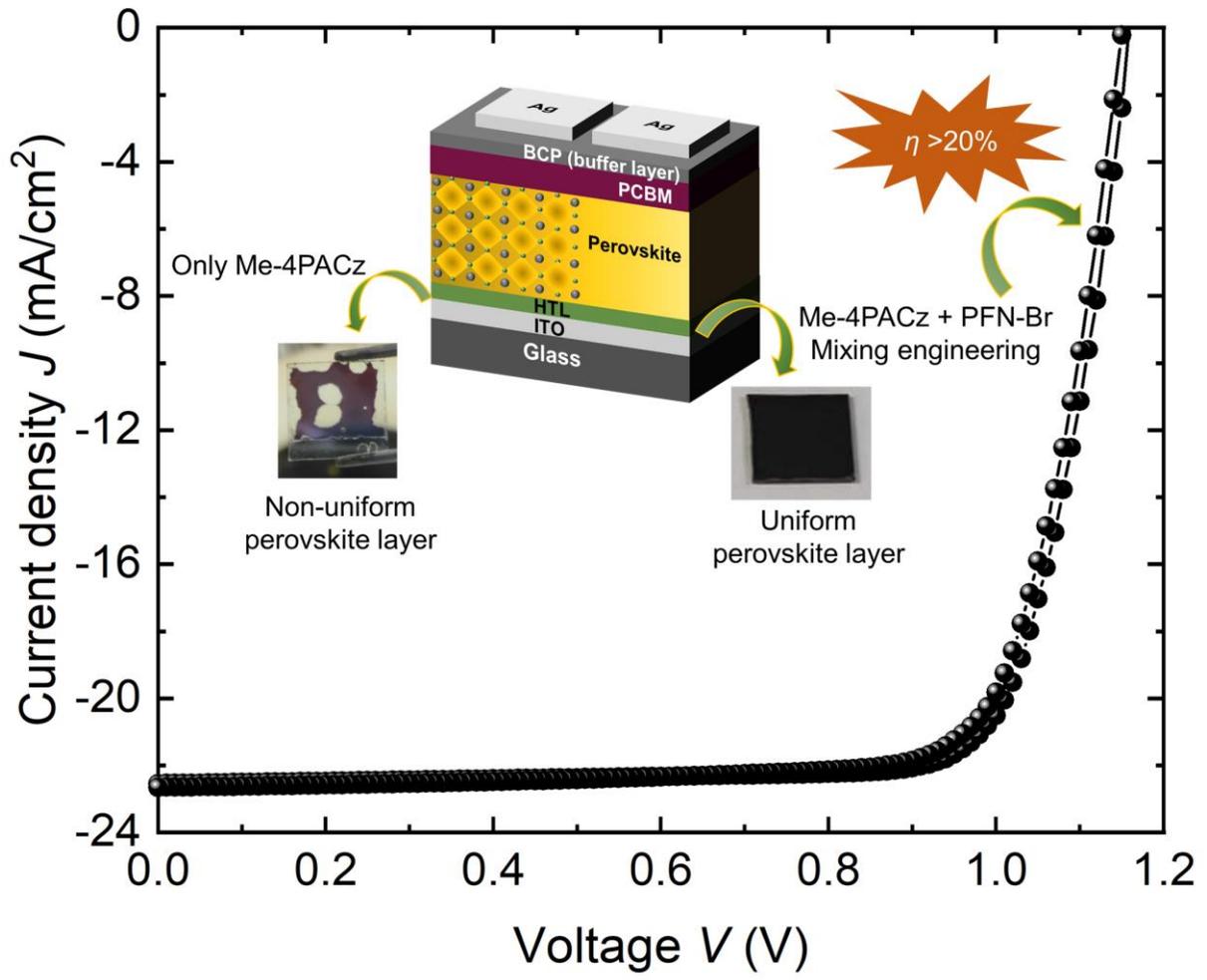